\documentstyle[multicol,aps,prl]{revtex}

\date{\today}
\draft
\tighten
\begin{document}
\def\sqr#1#2{{\vcenter{\hrule height.3pt
      \hbox{\vrule width.3pt height#2pt  \kern#1pt
         \vrule width.3pt}  \hrule height.3pt}}}
\def\square{\mathchoice{\sqr67\,}{\sqr67\,}\sqr{3}{3.5}\sqr{3}{3.5}}
\def\today{\ifcase\month\or
  January\or February\or March\or April\or May\or June\or July\or
  August\or September\or October\or November\or December\fi
  \space\number\day, \number\year}

\def\Bbb{\bf}


\newcommand{\ww}{\mbox{\tiny $\wedge$}}
\newcommand{\pp}{\partial}

\title{Gravitomagnetism in Brane-Worlds}
\author{Ali Nayeri~\thanks{email: nayeri@MIT.EDU} and 
        Adam Reynolds~\thanks{email: adam1@MIT.EDU}}
\address{\qquad\\
	Department of Physics,
	Massachusetts Institute of Technology \\
	77 Massachusetts Avenue,
	Cambridge, MA 02139, USA\\
	\date{\today}}
\maketitle

\begin{abstract}
In this paper we discuss a physical observable which is drastically
different in a brane-world scenario.  To date, the Randall-Sundrum
model seems to be consistent with all experimental tests of general 
relativity.
Specifically, we examine the so-called gravitomagnetic effect in the
context of the Randall-Sundrum (RS) model.  This treatment, of course,
assumes the recovery of the Kerr metric in brane-worlds which we have
found to the first order in the ratio of the brane
separation to the radius of the AdS$_5$, $(\ell/r)$.  We first show that 
the second Randall-Sundrum model of one brane leaves the gravitomagnetic 
effect unchanged.  Then, we consider the two-brane scenario of the original
Randall-Sundrum  proposal and show that the magnitude of the
gravitomagnetic effect depends heavily on the ratio of $(\ell/r)$.  
Such dependence is a result of the geometrodynamic spacetime and does not 
appear in static scenarios.  We hope that  we will be able to test this 
proposal experimentally with data from NASA's Gravity Probe B (GP-B)and possibly 
disprove either the Randall-Sundrum two-brane scenario or standard general 
relativity.

\vspace*{1mm}
\end{abstract}


\begin{multicols}{2}
\section{Introduction}

Motivated by earlier works on the extra dimensional solution to the hierarchy
problems of the fundamental interactions~\cite
{Arkani-Hamed:1998rs,Antoniadis:1998ig}, Randall and Sundrum 
have proposed a five-dimensional model of
the universe in which one can recover four-dimensional Einstein gravity at
large enough distances \cite{RS1}.  This model consists of a four-dimensional
hypersurface (3-brane) embedded in a five-dimensional Anti de Sitter
Spacetime (AdS$_5$).  The novelty of this theory is  that there is
no need for compactification of the extra dimension.  Instead, an
effective dimensional reduction occurs in the form of a massless
graviton localized near the brane in a `bound state.'  In the low
energy limit, this bound state dominates over the Kaluza-Klein (KK)
modes and all the gauge fields, which are constrained to live on the brane and
feels normal four-dimensional gravity.  Thus, as long as the length scale
of the AdS$_5$ is sufficiently small, standard Newtonian and general
relativistic results can be recovered to high accuracy~\cite{garr}.

A generic, non-cosmological form of the metric in this {\it non-factorizable}
spacetime has the form 
\begin{equation}
ds_5^2 = e^{-2|y|/\ell} g_{\mu \nu}dx^\mu dx^\nu + dy^2,
\label{RSmetric}
\end{equation}
where $\ell$ is the curvature radius of the AdS$_5$ and $g_{\mu \nu}$
must be asymptotically flat (for cosmological form and its generalization see,
e.g,~\cite{ali} and references therein). The extra dimension is denoted by $y$.
In other words, the induced metric is any
vacuum solution of the {\it four} dimensional Einstein's equations.

One example of this sort is the recovery of 
the 4-dimensional Schwarzschild metric for $g_{\mu \nu}$ to first
\cite{schwarz1} and second order in $(\ell/r)$ \cite{schwarz2}.  Here, the 
solution
is a black string hidden behind a ``cigar'' horizon.  The dependence
of the metric elements on the extra dimension makes the Einstein equations
extremely hard to solve exactly. 

However, it is because of this recovery that
people believe that the RS scenario is consistent with
all tests of general relativity.  We would like, however, to point out
that general relativity itself is  awaiting  for yet another test  to be
observed: ``{\it gravitomagnestism effect}''.  This is a pure relativistic
effect and there is no Newtonian counterpart.  It is mainly due to
the rotation of a massive body in the spacetime. While general relativity is 
being tested with the Gravity Probe B experiment, it is worthwhile to see whether 
the RS model is consistent 
with this yet to be tested effect of the general relativity.  If the
RS scenario is compatible with the result of Gravity
Probe B then this model is really consistent with the all general 
relativistic effects.

In order to discuss
topics related to rotation and frame-dragging, one has to
demonstrate the recovery of the Kerr metric.  This itself would
be a very difficult problem due to its lack of symmetry compare
with the Schwarzschild metric.  While in the case of spherical symmetry,
we have three unknown parameters to deal with, there are five
unknown parameters in the axial symmetry case. All parameters are
functions of two intrinsic spatial dimensions and the extra dimension.

In this paper, we begin with a brief review of the standard
gravitomagnetism and outline how it will be measured with Gravity
Probe B.  This is followed by a section describing the recovery of the
Kerr metric in the slow motion weak field limit and a section discussing 
gravitomagnetism in brane-worlds.
We conclude with our results and how they may be applied to
experimental data. Here and throughout the paper,  Greek indices indicate 
the usual four dimensions taking values 0-3 while Latin indices take values 
of 0-3 and 5 to include the extra dimension.  We shall work with the signature
$(-, +, +, +, +)$.

\section{Gravitomagnetism and its Measurement: A brief review}

Gravitomagnetism, also known as the Lense-Thirring effect, can be
understood with a close analog to classical electrodynamics.  As a
body of charge produces an electric field in electromagnetism, so a
body of mass creates a gravitational field in Newtonian gravity.  By
considering general relativity we can extend this analogy to include
magnetism.  Just as a current of charge produces a magnetic field, so
a matter current will give rise to a gravitomagnetic field (see
\cite{mash} and references therein).

For our purposes, we will be considering a slowly rotating, spherical
body such as the earth.  In this case, we need the Kerr Metric in the
slow-motion, weak-field limit.  This has the form (in Boyer-Lindquist
coordinates with the charge, $Q=0$),
\begin {eqnarray}
ds_4^2& = & -F(r) dt^2 + F^{-1}(r)dr^2 - {\frac{4G_4J}{r}}\sin ^2\theta 
          d\phi dt \nonumber \\
      &   & + r^2d\Omega ^2 ,\label{okerr}
\end {eqnarray}
where \( F(r)=(1- 2G_4Mr^{-1}) \) and $J$ is the
angular momentum of the spinning body. $G_4$ is the four dimensional
Newton's constant. We treat this as a linear
perturbation to Minkowski spacetime of the form (see~\cite{wheel}, for
instance),
$
h_{\mu \nu} \cong g_{\mu \nu} - \eta _{\mu \nu},
$
where $|h_{\mu \nu}|\ll 1$.

Once we perform an infinitesimal coordinate transformation and impose
the Lorentz gauge condition, ~$h^{\mu \nu}_{\ \ ,\nu}=0$ \cite{mash},
we can write the field equations as
\begin{equation}
\Box_4 h_{\mu \nu} = -16\pi G_4\left(T_{\mu \nu}-{1 \over 2}
\eta_{\mu \nu}T\right),
\end{equation} 
with \( \Box_4 \equiv \eta^{\mu \nu}{\frac{\partial ^2}{\partial 
x^\mu \partial x^\nu}} \),
and $T$ the trace of the energy-momentum tensor, $T^\mu_{\ \mu}$.
One can then consider the temporal, off-diagonal terms of the energy
momentum tensor of a non-relativistic matter which take the
form
\begin{equation}
\nabla ^2 h_{0i} \cong 16\pi G_4 \rho v_i.
\label{poisson1}
\end{equation}
This is easily recognizable as Poisson's equation.  In considering a
rotating charge in electromagnetism, one would find it useful to
define the magnetic dipole moment.  Here, we instead define the
angular momentum, ${\bf J}$, which can be thought of as an {\it
angular dipole moment.}  Then one can naturally obtain the
gravitomagnetic vector potential for a spinning mass
\begin{equation}
{\bf h }\cong -2G_4{\frac{\bf J\times {\bf\hat r}}{r^3}}.
\end{equation}
If we orient our body such that the ${\bf J}$ is along the $z$ axis, 
we have in spherical coordinates
$
h_{0\phi} \cong -(2G_4J\sin ^2\theta/r),
$
which is just the $g_{0\phi}$ component of the Kerr metric~(\ref{okerr}).  
Not surprisingly, we can also define the gravitomagnetic field
\begin{equation}
{\bf H}_{GR}{\bf \equiv \nabla \times h} \cong 2G_4 \left [ {\frac {\bf J - 
3({\bf J \cdot \hat r}){\bf \hat r}}{r^3}}\right ].
\end{equation}
With this formalism in place, one can consider the torque on a
gyroscope with angular momentum ${\bf S}$ due to ${\bf H}$,
\begin{equation}
{\bf \tau \cong} {\frac{1}{2}}{\bf S \times H} ={\frac{d\bf S}{dt} 
\equiv {\bf\dot\Omega \times  S}}.
\end{equation}
Therefore the gyroscope undergoes precession with angular velocity
\begin{equation}
{\bf \dot \Omega} = -{\frac{1}{2}}{\bf H} = -G_4 \left [ {\frac {{\bf J} - 
3({\bf J \cdot \hat r}){\bf \hat r}}{r^3}} \right ],
\end{equation}
which is very similar to the Larmor precession one would observe of a
magnetic moment precessing around a constant Magnetic field in
classical electromagnetism.

GP-B is a satellite experiment being designed and
implemented by NASA in conjunction with Stanford University
\cite{pugh:me1959,schiff:pr1960}. It is scheduled to be launched in the next
year or two \cite{gp-b} and it is hoped that it will provide the first
experimental test of the gravitomagnetic effect.  The satellite will
contain four gyroscopes made of fused quartz.  These gyros will be
electrically suspended and spinning in vacuum.  They will be coated in
niobium and cooled below the superconducting threshold with liquid
helium.  Superconducting Quantum Interference Devices (SQUID's) will
be used to measure the London moment which points along the spin axis
of the gyros.  The SQUID's will measure changes in the direction of
the axis as small as 0.1 milliarcseconds.  This system will be
implemented to measure the gyroscopic precession caused by the Earth's
gravitomagnetism over one year to 1\% or better.  Standard General
Relativity predicts that this will be 42 milliarcseconds \cite{wheel}.

\section{Recovery of the Kerr Metric in Brane-Worlds}

If the four-dimensional metric in (\ref{RSmetric}) is the Minkowski metric, 
$\eta_{\mu \nu}$, then the five-dimensional metric will satisfy the vacuum 
Einstein equations
\begin {equation}
R_{AB} - {\frac{1}{2}}Rg_{AB} - \Lambda g_{AB} = 0,
\label{5dEin}
\end {equation}  
with the bulk cosmological constant $\Lambda = -(6/\ell^2)$.
Due to the Randall-Sundrum fine tuning, these five-dimensional
Einstein equations with a bulk cosmological constant reduce to the
four-dimensional Einstein equations in vacuum without an effective
cosmological constant \cite{RS2}.  Thus we can rewrite (\ref{5dEin}) as
\begin{equation}
R_{\mu \nu} - 4\ell ^{-2}g_{\mu \nu} =0,\; \;
R_{5\mu} = 0,\; \; 
R_{55}-4\ell^{-2} = 0.
\label{4dEin}
\end{equation}
We can now relax our requirement on $g_{\mu \nu}$ and only
restrict it to being Ricci flat.  Now any ordinary, four-dimensional
solution to the vacuum Einstein equations can serve as our metric.

If we treat the Kerr metric in the slow motion and weak field limit
as a perturbation to the Minkowski background then following \cite{garr}
one can write the induced metric perturbation on the brane as the summation of
two parts: the part due to the matter fields on the brane and the part due to
the brane displacement, 
\begin{equation}
\bar h_{\mu\nu} = h^{(m)}_{\mu\nu} + 2 \ell^{-1}\gamma_{\mu\nu}\hat \xi^5
\label{hbrane},
\end{equation}
where $\hat \xi^5(x^{\mu}) = -y$ is the brane displacement due to the matter source
on the brane and $\gamma_{\mu\nu} = e^{-2|y|/\ell}\eta_{\mu\nu}$.  The matter
fields part $h^{(m)}_{\mu\nu}$ is obtained through following relation~\cite{garr},
\begin{equation}
h^{(m)}_{\mu\nu} = - 2 \kappa_5^2 \ \int d^4 x' G^{(R)}_5(x, x')
\left(T_{\mu\nu} - \frac{1}{3} \gamma_{\mu\nu}T\right)(x'),
\label{hm}
\end{equation}
where $\kappa_5^2 = 3 \pi^2 G_5$~\cite{mn}.  $T_{\mu\nu}$ is the energy-momentum
tensor of the brane matter fields and the contribution from the brane
tension has been excluded, though its effect has been considered in order to derive
(\ref{hm}).  Here $T = T^{\mu}_{\ \mu} = (6/\kappa^2_5)\Box_4 \hat \xi^5$
\cite{garr} and $G^{(R)}_5(x, x')$ is
the 5D retarded Green's function which satisfies
\end{multicols}
     \begin{eqnarray}
     & & \hspace{-3.63in} \line(246,0){246} \line(0,6){6} \nonumber
     \end{eqnarray}
     \begin{equation}
     \left[e^{+2|y|/\ell}\Box_4 + \partial^2_y - 4\ell^{-2}
     + 4\ell^{-1}\delta(y)\right]G^{(R)}_5(x, x') =
     \delta_5(x - x'),
     \label{eqm}
     \end{equation}
where the so-called RS gauge in which $h_{55} =
h_{\mu 5} = 0$, $h_{\mu \ \ ,\nu}^{\ \nu} = 0$ and $h^{\mu}_{\ \mu} = 0$
was used in arriving (\ref{eqm}).  Following ~\cite{RS1}, one can obtain 
the Green's function from a complete set of eigenstates
\begin{equation}
G^{(R)}_5 = - \int~\frac{d^4k}{(2\pi)^4}~e^{ik_{\mu}(x^{\mu} - 
x'^{\mu})}
\left[\frac{\ell^{-1}e^{-2(|y| + |y'|)/\ell}}{{\bf k}^2 - 
(\omega + 
i\epsilon)^2}
+ \int_0^{\infty}~dm\frac{u_m(y)u_m(y')}{m^2 + {\bf k}^2 - 
(\omega + i\epsilon)^2}
\right],
\end{equation} 
     \begin{eqnarray}
     & & \hspace{3.67in} \line(0,-6){6} \line(245,0){245} \nonumber
     \end{eqnarray}
\begin{multicols}{2}
\noindent where the first term corresponds to zero mode contributions and 
the second term
shows the contributions from the continuum KK modes $u_m(y) = \sqrt{m\ell/2}
\{J_1(m\ell)Y_2(m\ell e^{|y|/\ell}) - Y_1(m\ell)J_2(m\ell e^{|y|/\ell})\}
/\sqrt{J_1(m\ell)^2 + Y_1(m\ell)^2}$.  Now in the case of the Kerr metric,
the energy-momentum tensor can be written as $T_{\mu\nu} = \rho(r) v_{\mu} v_{\nu}$
and since we are dealing with the stationary, axial symmetry case by using 
Eq.~(\ref{hbrane})
we find that
\begin{eqnarray}
\bar h_{00} & = & \frac{2G_4M}{r}\left( 1 + \frac{2}{3}\frac{\ell^2}{r^2} +...
\right), \nonumber \\
\bar h_{0i} & = & - 2 A_i\left( 1 + \frac{3}{2}\frac{\ell^2}{r^2} + ...
\right), \\
\bar h_{ij} & = & \frac{2G_4M}{r}\left( 1 + \frac{1}{3}\frac{\ell^2}{r^2} +...
\right)\delta_{ij}, 
\label{hkerr}
\nonumber
\end{eqnarray}
where $r = |{\bf x} - {\bf x'}|$, $G_4 = (3\pi G_5/ 8\ell)$ is
the four-dimensional Newton's constant, and $M = \int~d^3x\rho$
is the total mass. ${\bf A}$ is the gravitomagnetic vector potential
which has the form ${\bf A} \sim G_4 ({\bf J} \times {\bf r})/r^3$.
In deriving (\ref{hkerr}) we have assumed that $|{\bf x}| > |{\bf x'}|$
and both points are on the wall ($y = y' = 0$).  Thus in the stationary case
\begin{eqnarray}
G^{(R)}_5({\bf x}, 0, {\bf x'}, 0) & = & \int_{-\infty}^{+\infty}
dt G^{(R)}_5(x, x') \nonumber \\
& \approx &  -\frac{1}{4\pi \ell r}\left[1 + \frac{1}{2}
\frac{\ell^2}{r^2} +
 ...\right].
\end{eqnarray}

Now, having all the components of the metric, it seems that up to the
first order in $(\ell/r)$, we recovered the Kerr metric (\ref{okerr})
in the slow motion and weak field limit on the brane. We would like,
however, to point out that as the weak field limit of the RS model, in
general, differs from the weak field limit of the normal four dimensional
Schwarzschild solution, the ratio of gravitomagnetic to gravitoelectric
metric coefficients is not the same as the one in the usual four dimensional
Kerr solution.  This means that the gravitomagnetic effect of the two models
will be different.

Though we recovered the Kerr metric in the slow motion and weak field
limit, one can obtain the general form of the Kerr metric by inserting 
the most general axially symmetric, (non-singular) stationary metric 
in four-dimensions into (\ref{RSmetric}) to obtain~\cite{later}
\end{multicols}
     \begin{eqnarray}
     & & \hspace{-3.63in} \line(246,0){246} \line(0,6){6} \nonumber
     \end{eqnarray}
\begin{equation}
\nonumber ds_5^2  =  e^{-2|y|/\ell}\left[-e^{2\nu}dt^2 + e^{2\psi}
(d\phi-\omega dt)^2 + 
e^{2\mu_2}(dx_2)^2 + e^{2\mu_3}(dx_3)^2\right] + dy^2,
\label{4dmetric}
\end{equation}
where $\nu, \psi, \omega, \mu_2$ and $\mu_3$ are functions of
$x_2$,$x_3$ and $y$.  There is an unexploited gauge freedom between
$\mu_2$ and $\mu_3$ \cite{chand}.  This metric can then be substituted
into (\ref{4dEin}) to yield a series of coupled partial differential
equations in $x_2, x_3$ and $y$.  After rigorous calculations,
one can show, for sufficiently small $\ell$ that
calculations that
\begin{equation}
ds_5^2  \approx  e^{-2|y|/\ell}
\left[- \rho^2\frac{\Delta}{\Sigma^2} dt^2
+ \frac{\Sigma^2}{\rho^2}\left(d\varphi + \frac{2G_4Jr}{\Sigma^2}dt\right)^2
\sin^2{\theta}
  + \frac{\rho^2}{\Delta}dr^2 + \rho^2 d\theta^2\right ]
+ dy^2 \label{bkerr},
\end{equation}
     \begin{eqnarray}
     & & \hspace{3.67in} \line(0,-6){6} \line(245,0){245} \nonumber
     \end{eqnarray}
\begin{multicols}{2}
\noindent where $\Delta (r, \theta, 0) = r^2 - 2 G_4 Mr + (J/M)^2$ 
and $\rho^2 = r^2 + 
(J/M)^2\cos^2{\theta}$ with 
$
\Sigma^2 = \rho^2\left(r^2 + a^2\right) + 
2(J/M)^2G_4Mr\sin^2{\theta}. 
$
In arriving at (\ref{bkerr}), the gauge, $e^{(\mu_3 - \mu_2)} =
\sqrt{\Delta (x_2, x_3, y)}$ has been chosen.
The details of these calculations are the subject of
our upcoming paper \cite{later}.  

\section{Gravitomagnetism in Brane Worlds}
With the Kerr metric recovered in the slow motion and weak field
limit, we can now investigate gravitomagnetism in the context of a
brane-world scenario.  It has been shown \cite{garr} that one can
obtain the following field equations for one brane of positive
tension:
\begin{equation}
\Box_4 h_{\mu\nu} = -16 \pi G_4 \left(T_{\mu\nu}-{1\over 2}
\gamma_{\mu\nu} T \right),
\end{equation}
with the brane sitting at $y = 0$.  
We can then easily write down the corresponding version of 
(\ref{poisson1}) as
$
\nabla^2 h_{0i} \cong -16 \pi G_4 \rho v_i,
$
for the brane at $y=0$, and we recover the normal gravitomagnetic
potential without a correction.  This is no surprise as the RS single
brane scenario is effectively four dimensional field theory. However, 
in the case of two branes, one of positive tension (shadow brane)at $y=0$ 
and one of negative tension (our observable universe) at $y=d$, the situation
will be different as one can observe the effect of shadow matter on our universe.  
In this case, we find Einstein equations
of the form \cite{garr}
\begin{eqnarray}
\nonumber \left(e^{2|y|/\ell}\Box h_{\mu\nu}\right)^{(\pm)}
= -\sum_{\sigma=\pm} 16\pi G^{(\sigma)} \left(T_{\mu\nu}-{1\over 3}
  \gamma_{\mu\nu} T\right)^{(\sigma)}   \\
\pm {16\pi G^{(\pm)}\over 3}{\sinh(d/\ell)\over e^{\pm d/\ell}} 
\gamma_{\mu\nu} T^{(\pm)},
\end{eqnarray}
with $ G^{(\pm)}=[3\pi G_5\ell^{-1} e^{\pm d/\ell}/ 16 \sinh(d/l)]. $
For the brane with negative tension, where our universe resides, this yields
\begin{equation}
e^{2d/\ell}\nabla^2h_{0i} \cong 16\pi G_4\rho v_i,
\end{equation}
where we have used the fact that the four-dimensional Newton's
gravitational constant satisfies the relation $G_4 = 3\pi G_5/8
\ell\tanh(d/\ell)$ for the same matter distributions on two branes.

Our correction to the gravitomagnetic potential is similarly
manifested in gravitomagnetic field, ${\bf H}_{GR}$:
\begin{equation}
{\bf H}_{RS} = {\bf H}_{GR}e^{-2d/\ell}
\end{equation}
To solve the hierarchy problem, the warp factor must bridge a
gap of sixteen orders of magnitude.  This indicates that $d/\ell \sim
36$ and gravitomagnetism is also decreased by $32$ orders of
magnitude from the standard general relativistic prediction.  This
puts the gravitomagnetic field of the Earth far outside the reach of
GP-B.  Thus, this Randall-Sundrum model predicts that Gravity Probe B
will see {\it no} gravitomagnetic field.  If, however, a nonzero
precession is measured, then the two-brane scenario must be amended.
On the other hand, if GP-B will be able to measure the GR predications
then this will put an upper bound on $(d/\ell)$ of $1.15$ which is in sharp
contrast to what you may find by observing the Brans-Dicke bound.

At the end we again would like to emphasize that GP-B is not merely another
test of GR and could serve to test models with extra dimensions such
as the RS model.  Another point which we would like to bring to attention 
is that
it is not sufficient for a theory of higher dimensions to just satisfy
the Schwarzschild type of experimental test and it seems that the Kerr test
provide deeper understanding of such models. 

\section*{Acknowledgments} 
The authors would like to thank E. Bertschinger, A. Chamblin, and A. Guth
for insightful discussion and useful comments. A.N. is supported by NSF
grant ACI-9619019.

\end{multicols} 
\end{document}